 \documentclass[aps,pra,superscriptaddress,amsmath,amssymb,preprintnumbers,twocolumn,floatfix,showpacs,showkeys,10pt]{revtex4-1}
 \usepackage{amssymb} \usepackage{epsfig}
 \begin{document}
  \title{Witnessing criticality in non-Hermitian systems via entropic uncertainty relation}
\author{You-neng Guo}
\email{guoxuyan2007@163.com}
\affiliation{Department of Electronic and Communication Engineering, Changsha University, Changsha, Hunan
410022, People's Republic of China}
\author{Guo-you Wang}
\email{gywang@hut.edu.cn}
\affiliation{College of Science, Hunan University of Technology, Zhuzhou 412008, People's Republic of China}

\begin{abstract}
Non-Hermitian systems with exceptional points lead to many intriguing phenomena due to the coalescence of both eigenvalues and corresponding eigenvectors, in comparison to Hermitian systems where only eigenvalues degenerate. In this paper, we have investigated entropic uncertainty relation (EUR) in a non-Hermitian system and revealed a general connection between the EUR and the exceptional points of non-Hermitian system. Compared to the unitary dynamics determined by a Hermitian Hamiltonian, the behaviors of EUR can be well defined in two different ways depending on whether the system is located in unbroken or broken phase regimes. In unbroken phase regime, EUR undergoes an oscillatory behavior while in broken phase regime where the oscillation of EUR breaks down. The exceptional points mark the oscillatory and non-oscillatory behaviors of the EUR. In the dynamical limit, we have identified the witness of critical behavior of non-Hermitian systems in terms of the EUR. Our  results reveal that the EUR witness can exactly detect the critical points of non-Hermitian systems beyond (anti-) PT-symmetric systems. Our results may have potential applications to witness and detect phase transition in non-Hermitian systems.

 \end{abstract}

 \maketitle
\section{Introduction}
Historically, the uncertainty principle $\Delta x \Delta p \geq \hbar/2$, originally proposed by Heisenberg~\cite{Heisenberg} in 1927, captures that one can't simultaneously predict the measurement outcomes of position and momentum with certainty. Subsequently, it has been further formulated by Robertson~\cite{Robertson} and generalized to arbitrary pairs of incompatible observables $R$ and $Q$: $\Delta R\cdot\Delta Q\geq\frac{1}{2}|\langle\psi| [R,Q] |\psi\rangle|$, here $\Delta R(\Delta Q)$ denotes the standard deviations and $[R,Q]=RQ-QR$ is the commutator. However, this uncertainty relation in terms of the commutator is not always optimal to quantify the measuring uncertainty, because the bound is strongly related to the state $|\psi\rangle$ of system. In particular, there is a trivial result when $\langle\psi| [R,Q] |\psi\rangle$ is equal to zero for some specific states even though $R$ and $Q$ do not share any common eigenvectors. To overcome this
drawback, Deutsch~\cite{Deutsch} in 1983 originally developed the so-called entropy uncertainty relation (EUR) in an information theoretical framework instead of the standard deviation, which later was proven by Maassen and Uffink~\cite{Maassen} in 1988
\begin{equation}\label{Eq1}
H(R)+H(Q)\geq -2\log_{2}c
\end{equation}
where $H(R)$($H(Q))$ denotes the Shannon entropy of the probability distribution of the outcomes when $R$($Q$) is performed on the pure state $|\psi\rangle$ of system, respectively. $c=\max_{i,j}|\langle\varphi_{i}|\psi_{j}\rangle|$ quantifies the complementarity of $R$ and $Q$ with their corresponding eigenvectors $|\varphi_{i}\rangle$ and $|\psi_{j}\rangle$. However, for any mixed state $\rho$ of system,  the EUR can be formulated in
general form as
\begin{equation}\label{Eq2}
H(R)+H(Q)\geq -2\log_{2}c + S(\rho),
\end{equation}
here $S(\rho)=-Tr(\rho \log_{2}\rho)$ is the von Neumann entropy.

A distinct advantage of the EUR superior to the standard deviations is that the lower bound of EUR does not depend on specific states to be measured. In fact, the EUR can be further improved with the assistance of a quantum memory, so that the outcomes of two incompatible measurements can be predicted precisely by an observer with access to the quantum memory if the initial states are maximally entangled~\cite{Renes,Boileau}. At present, the EUR has received a great deal of attention~\cite{Berta,Prevedel,Li,Adabi,Adabi1,ZhangY1,Chen,Chen1,Huang1,ZhangY2,Huang3,Huang4} due to potential applications in quantum information processing tasks such as quantum entanglement witnessing~\cite{Hu,Pati,Xu,Xu1}, and quantum key distribution~\cite{Hu1,Pourkarimi}.

Unfortunately, previous efforts have attempted to discuss the EUR in Hermitian systems. Few detailed investigations concerning the EUR in non-Hermitian systems are available. In the present work, we extend the EUR dynamics of Hermitian systems to non-Hermitian ones. The primary motivation for such an extension is at least two-fold. First, there has been a growing interest in non-Hermitian systems due to the fact that many intriguing phenomena inaccessible in Hermitian systems can be observed in non-Hermitian systems, e.g., the violation of no-signaling principle~\cite{Lin,Regensburger}, loss-induced lasing~\cite{Brandstetter,Peng}, and the optimal brachistochrone problem~\cite{Huang0} and so on. Therefore, it is natural to ask whether the EUR can display some new phenomena in the non-Hermitian system. Second, in contract to Hermitian systems, there exists so-called exceptional points in non-Hermitian systems where not only eigenvalues, but also their corresponding eigenvectors coalesce. It has been proved that such points are relevant to describe dynamical phase transitions from an unbroken phase to a spontaneous broken phase~\cite{Yang, Miri}. Therefore, it is urgently to wander whether this critical phenomena of non-Hermitian system can be confirmed by the behaviors of EUR.

Motivated by the above issues, we first focus on the behaviors of EUR in non-Hermitian systems and reveal a general connection between the EUR and the critical points of non-Hermitian system. Compared with the dynamics of EUR in Hermitian system case, we find that the dynamics of EUR can be well defined in the unbroken phase to the broken phase regimes. In the unbroken phase regime, the EUR exhibits an oscillatory behavior, while for the broken phase regime, the oscillation of the EUR breaks down. In this respect, this provides us a powerful method to witness criticality in non-Hermitian systems. Particularly, we identify unique criticality based on the EUR in the time limit infinity around the critical point, above which the EUR witness increases asymptotically but below which the EUR witness decays asymptotically. Therefore, there exists a sudden change of the EUR witness at critical points which are referred to the exceptional points of non-Hermitian system.

\section{Dynamics of a general two-level non-Hermitian system}\label{model}

In order to demonstrate the behaviors of EUR in a non-Hermitian system, for clarity and without loss of generality, we take a general two-level non-Hermitian system whose Hamiltonian reads as~\cite{Benderc}
\begin{eqnarray}\label{eq:H1}
	\mathcal{H}_{NH} =\begin{pmatrix}
	             r e^{i\phi}          &         \sigma   \\
	              s           &        r e^{-i\phi}
	              \end{pmatrix}.
	\end{eqnarray}
The energy eigenvalues of $\mathcal{H}_{NH}$ are $ E_{\pm} = r\cos\phi\pm \sqrt{s\sigma - r^2\sin^{2}\phi}$ and the corresponding eigenvectors are
$|E_{\pm}\rangle=\frac{1}{\sqrt{2\cos \Theta}}\left(
\begin{array}{ c c l r }
 e^{\pm i \Theta/2} \\
\pm e^{\mp i \Theta/2}  \\
\end{array}
\right)$, respectively.
Here, defining $\sin\Theta =r\sin\phi/\sqrt{s\sigma}$. Obviously, two regions that are separated by the point $s\sigma = r^2\sin^{2}\phi$ where the phase transition from an unbroken phase to a  broken phase happens. For $ s\sigma > r^2\sin^{2}\phi $, the system is denoted in unbroken phase regime and exhibits real spectra, while for $ s\sigma < r^2\sin^{2}\phi $, the system is termed in broken phase regime where complex conjugate eigenvalues emerge.

On the other hand, for the case of $s=\sigma$, Eq. (\ref{eq:H1}) reduces to $PT$ symmetric Hamiltonian which is invariant under the $PT$ transformation $(PT)H(PT)^{-1}=H$ with operator $P$ denotes Pauli matrix $\sigma_{x}$, and $T$ represents the complex conjugation~\cite{Bender0}. The corresponding eigenvalues are purely real in the unbroken phase regime $ (s > |r\sin\phi|) $ and become complex in the broken phase regime$ (s < |r\sin\phi|)$~\cite{Guo}.
	
Using the non-Hermitian Hamiltonian of Eq. (\ref{eq:H1}), the dynamics of the non-Hermitian system can easily be obtained by solving the time dependent Schr\"{o}dinger equation
	\begin{equation}\label{eq:H2}
	i\frac{\partial |\Psi (t)\rangle}{\partial t} = \mathcal{H}_{NH} |\Psi (t)\rangle,
	\end{equation}
with the non-unitary time evolution operator $ \mathcal{U}(t)=\exp[-i\mathcal{H}_{NH} t]$ which is obtained as,
\begin{equation}\label{non-unitary}
    \mathcal{U}(t) = e^{-itr\cos\phi}\begin{pmatrix}
\cos\omega t + \frac{r\sin\phi\sin \omega t}{\omega}              &          -i \frac{\sigma\sin\omega t}{\omega} \\\\
	              -i \frac{s\sin\omega t}{\omega}               &         \cos\omega t - \frac{r\sin\phi\sin \omega t}{\omega}
	              \end{pmatrix},
	\end{equation}
where $\omega = \sqrt{s\sigma - r^2\sin^{2}\phi}$. Particularly, at the exceptional point, namely, $ s\sigma = r^2\sin^{2}\phi$, the time evolution operator becomes

\begin{equation}\label{eq:H7}
    \mathcal{U}(t) \approx e^{-itr\cos\phi}\begin{pmatrix}
	              1 + tr\sin\phi               &          -i tr\sin\phi \\\\
	              -i tr\sin\phi              &          1 - tr\sin\phi
	              \end{pmatrix}.
	\end{equation}
It is worthwhile to mention that, the non-Hermitian Hamiltonian given by Eq. (\ref{eq:H1}) with real spectra can be mapped into
a Hermitian one via the Hermitian transformation~\cite{Mostafazadeh,Zhou,Zhou0}, e.g., $\eta \mathcal{H}_{NH} \eta^{-1}=H$, where the Hermitian matrix is $\eta=\frac{1}{\sqrt{\cos \Theta}}\left(
\begin{array}{ c c c c l r }
 \cos(\Theta/2)  & -i \sin(\Theta/2)\\
i \sin(\Theta/2)  & \cos(\Theta/2)  \\
\end{array}
\right)$. Therefore, the non-unitary time evolution operator given by Eq. (\ref{non-unitary}) reduces to an unitary evolution one as a consequence of Hermiticity of $H$
\begin{equation}\label{eq:H8}
    \mathcal{U}(t) = e^{-itr\cos\phi}\begin{pmatrix}
	              \cos\omega t                 &          -i \sin\omega t \\\\
	             -i \sin\omega t               &          \cos\omega t
	              \end{pmatrix}.
	\end{equation}
Given an arbitrary initial state $|\Psi(0)\rangle$, one can express it as a superposition of $|E_{\pm}\rangle$, namely, $|\Psi(0)\rangle=\sin\frac{\theta}{2}|E_{+}\rangle+\cos\frac{\theta}{2}|E_{-}\rangle$ which evolves in time according to $|\Psi(t)\rangle =\mathcal{U}(t)|\Psi(0)\rangle=\sin\frac{\theta}{2} e^{-iE_{+}t}|E_{+}\rangle+\cos\frac{\theta}{2} e^{-iE_{-}t}|E_{-}\rangle$. However, due to the nonorthogonality of eigenvectors $|E_{\pm}\rangle$, we have to transform this eigenvector representation into the computational orthonormal vectors $\{|0\rangle,|1\rangle\}$ via the similarity transformation
\begin{subequations}
\begin{align}
|0\rangle &  =\frac{1}{\sqrt{2\cos \Theta}}\left(e^{i \Theta/2}|E_{+}\rangle+e^{-i \Theta/2}|E_{-}\rangle\right)%
\label{Mapl1}\\
|1\rangle &  =\frac{1}{\sqrt{2\cos \Theta}}\left(e^{-i \Theta/2}|E_{+}\rangle-e^{i \Theta/2}|E_{-}\rangle\right).\label{Mapl2}%
\end{align}
\end{subequations}

In this respect, the time evolution state $|\Psi(t)\rangle$ can be rewritten as
\begin{eqnarray}\label{eq:H3}
|\Psi(t)\rangle &=&\left(\alpha e^{-iE_{+}t}-e^{-i\Theta}\beta e^{-iE_{-}t}\right)|0\rangle\\ \nonumber
&+&\left(e^{-i\Theta}\alpha e^{-iE_{+}t}+\beta e^{-iE_{-}t}\right)|1\rangle,
\end{eqnarray}
where $\alpha=\frac{1}{2}\sec\Theta(\cos\frac{\theta}{2}+e^{i\Theta}\sin\frac{\theta}{2})$ and $\beta=\frac{1}{2}\sec\Theta(e^{i\Theta}\cos\frac{\theta}{2}-\sin\frac{\theta}{2})$.
In general, the evolution of a non-Hermitian system is not trace-preserving. Therefore, a normalized form should be given
\begin{equation}
\rho(t)=\frac{|\Psi (t)\rangle\langle\Psi (t)|}{\text{tr}[\langle\Psi (t)|\Psi (t)\rangle]}.  \label{renorm}
\end{equation}

\section{Entropic uncertainty relation in non-Hermitian systems}\label{model}

To make calculations of the EUR $=-\Sigma_{X}p_{X}^{i}\log_{2}{p_{X}^{i}}$ which is only depending on the outcomes of probability, some measuring process is necessary to carry out. For simplify, we examine a pair of projective operators as observables $X \in (R,Q)$  which are represented by $P_{X}=1/2(I+\vec{n}\cdot\vec{\sigma})$, where $\vec{n}=(n_{1},n_{2},n_{3})$ is a unit vector and $\vec{\sigma}=(\sigma_{x},\sigma_{y},\sigma_{z})$. Therefore, the probability distribution of measurement outcome becomes
\begin{eqnarray}\label{Eqs3}
p_{X}&=&\frac{1}{N}\{2\alpha\beta^{\ast}[n_{3}\cos\Theta-i(n_{2}-\sin\Theta)]\\ \nonumber
&+&2\alpha^{\ast}\beta[n_{3}\cos\Theta+i(n_{2}-\sin\Theta)]e^{2i\Delta E t}\\ \nonumber
&+&[(n_{2}\sin\Theta-1)\sec^2\Theta-n_{1}\sin\theta]e^{i\Delta E t}\},
\end{eqnarray}
where $N=4i\sin\Theta(\alpha^{\ast}\beta-\alpha^{\ast}\beta e^{2i\Delta E t})-2\sec^2\Theta e^{i\Delta E t}$. From above expression, the outcome probability is associated with the difference of energy eigenvalues $\Delta E \equiv E_{+}-E_{-}=2\omega$ and leads to two completely different dynamical behaviors. For $ s\sigma > r^2\sin^{2}\phi $, the probability distribution exhibits an oscillatory behavior with the period $T=\pi/\omega$. In contrary, for $ s\sigma < r^2\sin^{2}\phi $, the probability distribution with oscillation breaks down. Particularly, for $ s\sigma = r^2\sin^{2}\phi $ which is referring to the exceptional point where the eigenvalues as well as their corresponding eigenvectors coalesce, and the probability distribution becomes
\begin{eqnarray}\label{Eqs4}
p_{X}&=&\frac{1}{4}[2-2n_{2}\\ \nonumber
&+&\frac{2(n_{2}+n_{3}\cos\theta+n_{1}\sin\theta+2n_{3}rt\sin\phi)}{1+r^2t^2+rt(2\cos\theta\sin\phi-rt\cos2\phi)}].
\end{eqnarray}

To reveal a relationship between the EUR and the exceptional point of non-Hermitian system, let us consider the initial state to prepare a maximal coherence pure state,
$
|\Psi\rangle=\frac{1}{\sqrt{2}}(|0\rangle+|1\rangle)
$
at time $t=0$. According to the Eq. (\ref{renorm}), a normalized density at a later time $t$ is
\begin{widetext}
\begin{eqnarray}\label{eq:Heff}
	\rho(t) = \begin{pmatrix}
	              \frac{\sigma^2\sin^2\omega t+(\omega\cos\omega t + r\sin\phi\sin \omega t)^2}{2T\omega^2}              &          \frac{[\omega\cos\omega t+(is - r\sin\phi)\sin\omega t][\omega\cos\omega t-(i\sigma -r\sin\phi)\sin\omega t]}{2T\omega^2}   \\\\
	               \frac{[\omega\cos\omega t-(is +r\sin\phi)\sin^2\omega t][\omega\cos\omega t+(i\sigma +r\sin\phi)\sin\omega t]}{2T\omega^2}         &      \frac{s^2\sin^2\omega t+(\omega\cos\omega t - r\sin\phi\sin \omega t)^2}{2T\omega^2}
	              \end{pmatrix},
	\end{eqnarray}
\end{widetext}
where $T=\cos^2\omega t + \frac{(s^2+\sigma^2+2r^2\sin^2\phi)\sin^2\omega t}{2\omega^2}$.  Considering the two
complementary observables $P=\sigma_{x}$ and $P=\sigma_{z}$ as measured observables (i.e., choosing $\vec{n} = (\pm1,0,0)$ and $\vec{n} = (0,0,\pm1)$ for $\sigma_{x}$ and $\sigma_{z}$ respectively), the lower bound of the EUR in Eq.(1) is always 1, while its EUR takes the form as
\begin{equation}\label{Eq11}
EUR=-\sum_{i=1}^{2}p_{x}^{i}\log_{2}{p_{x}^{i}}-\sum_{i=1}^{2}p_{z}^{i}\log_{2}{p_{z}^{i}}
\end{equation}
with

$p_{x}^{1}=\frac{4\omega^2\cos^2\omega t+(s+\sigma)^2\sin^2\omega t}{4T\omega^2}$,

$p_{x}^{2}=\frac{[4r^2\sin^2\phi+(s-\sigma)^2]\sin^2\omega t}{4T\omega^2}$,

$p_{z}^{1}=\frac{s^2\sin^2\omega t+(\omega\cos\omega t-r\sin\phi\sin\omega t)^2}{2T\omega^2}$,

$p_{z}^{2}=\frac{\sigma^2\sin^2\omega t+(\omega\cos\omega t+r\sin\phi\sin\omega t)^2}{2T\omega^2}$.
\begin{figure}[htpb]
  \begin{center}
     \includegraphics[width=7.5cm]{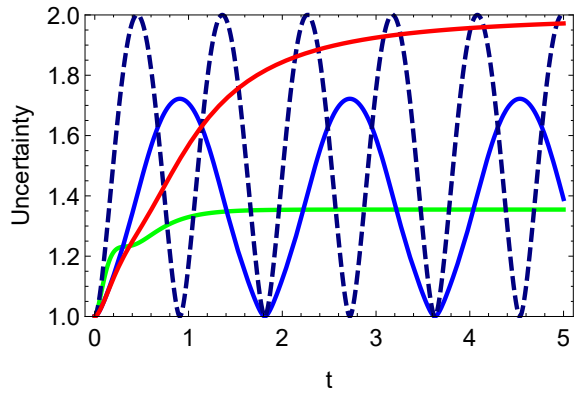}
     \caption{\label{fig1} (Color online) The EUR for initial state $|\Psi\rangle=\frac{1}{\sqrt{2}}(|0\rangle+|1\rangle)$ with observables $Q=\sigma_{x}$ and $R=\sigma_{z}$ with respect to $t$ in both Hermitian and non-Hermitian systems. Blue dashed curve depicts the behavior of EUR under the Hermitian Hamiltonian. Blue solid, green solid and red solid curves correspond to $ s\sigma > r^2\sin^{2}\phi$ (e.g.,$s=\sigma=2$ and $r=1$), $ s\sigma < r^2\sin^{2}\phi$ (e.g.,$s=\sigma=1$ and $r=2$) and exceptional points $ s\sigma = r^2\sin^{2}\phi$, respectively. Other parameters are fixed $\phi=\pi/2$. }
  \end{center}
\end{figure}

For comparison, we have also considered the case when time evolution is generated by a Hermitian Hamiltonian, whose expression of EUR can also be obtained by the similar procedure. Fig. 1 numerically shows the dynamics of EUR for initial state $|\Psi\rangle=\frac{1}{\sqrt{2}}(|0\rangle+|1\rangle)$ with observables $Q=\sigma_{x}$ and $R=\sigma_{z}$ in both Hermitian and non-Hermitian cases. Obviously, the behavior of EUR in a Hermitian system displays Rabi-type oscillation with the period $2\pi/\omega$ due to intrinsically unitary dynamical evolution. However, things become interesting for the non-Hermitian system. There are three distinct behaviors relating to the exceptional point $ s\sigma = r^2\sin^{2}\phi$. More specifically, for $ s\sigma > r^2\sin^{2}\phi$, the system in unbroken phase has pure real eigenvalues, and the EUR undergoes an periodic oscillatory behavior with $T=\pi/\omega$. This result is similar to what we have observed in the Hermitian case, since the non-Hermitian system with real eigenvalues is equivalent to Hermitian system by a similarity transformation~\cite{Gardas,Du}. While for $ s\sigma < r^2\sin^{2}\phi$ in broken phase regime, the system with complex spectra turns up, the oscillation of EUR breaks down. However, at the exceptional point, a phase transition occurs from an unbroken phase to a broken phase regime, the behavior of EUR increases asymptotically to a stable value. Therefore, the EUR in a non-Hermitian system is well defined by the exceptional point changing from oscillations to oscillations broken.

It is worthy mentioning that, the oscillatory or revival dynamical behavior of system is usually as an indicator of non-Markovian dynamics leading to information retrieval from the environment to the system~\cite{Kawabata,Cao}. Therefore, the physical origin behind oscillatory or revival of EUR in the unbroken phase is attributed to the backflow of information between the environment and the system, while monotone increasing EUR implies unidirectional information flow from the system to the environment. From the measure of uncertainty view, our results also indicate, one can predict measurement outcome more accurately in the unbroken regime compared to the broken regime due to the fact that the information retrieval from the environment can be retrieved in the unbroken phase, but not in the broken phase~\cite{Kawabata,Cao}.

\section{Entropic uncertainty relation as a signature of the exceptional points of non-Hermitian systems }

In traditional quantum information approaches, quantum phase transitions can be identified by the critical behavior of quantities such as entanglement~\cite{Gu,Radgohar,Serbyn}, quantum Fisher information~\cite{Prosen,Lambert}, fidelity~\cite{Cozzini,Sacramento,Panahiyan} and so on. These quantities are sensitive to the quantum criticality, and often serve as a signature of transition points. However, these quantities are more than a mere theoretical construct, especially they have difficult in experimental accessibility. Therefore, it is interesting to investigate the quantum criticality beyond Hermitian systems, and with a criticality witness that is feasible and experimentally accessible.

Recall that quantum phase transitions have their roots in purely quantum fluctuations due to the Heisenberg uncertainty relation, we identify a criticality witness based on the long-time average of the EUR
\begin{equation}\label{eq:H12}
	\mathcal{W} = \lim_{T\rightarrow\infty}\frac{1}{T}\int_{0}^{\infty}EUR(t)dt.
	\end{equation}
In analogy to what has been performed for quantum entanglement as a signature of the quantum criticality~\cite{Serbyn}, a sudden change of $\mathcal{W}$ with respect to varying system parameters occurs at the critical point implying that there is a quantum phase transition occurs, and this critical point
is referred to as an exceptional point in non-Hermitian systems.

\subsection{Criticality witness in PT-symmetric non-Hermitian systems}

In order to check the efficiency of the proposed criticality witness $\mathcal{W}$, we first consider the non-Hermitian systems with Hamiltonian given by Eq.(3). A quantum phase transition occurs at the exceptional point $ r_{0}=\sqrt{s\sigma/\sin^{2}\phi} $. In particular, for the case of $s=\sigma$, the system with Eq.(3) reduces to a standard PT-symmetric case, and a quantum phase transition occurs at $ r_{0}=s/\sin\phi $. To test the validity of the above mentioned criticality witness $\mathcal{W}$, we resort to numerical simulation.
Fig. 2(a) displays the behaviors of criticality witness $\mathcal{W}$ for non-Hermitian system with PT symmetry for $s=\sigma=2$ and $\phi=\pi/2$. As expected, the behaviors of criticality witness $\mathcal{W}$ increases asymptotically in the unbroken phase $r<r_{0}=2$, while in the broken phase $r>r_{0}=2$, the $\mathcal{W}$ displays asymptotically decay. A abrupt change of criticality witness $\mathcal{W}$ happens at the transition point $r=r_{0}=2$. This result implies the signature of a quantum critical point is confirmed by $\mathcal{W}$. On the other hand, the similar behavior occurs for a general non-Hermitian system without PT symmetry, e.g., $\sigma=\sqrt{2}$, $s=\sqrt{2}/2$ and $\phi=\pi/2$ as shown in Fig. 2(b). A sudden change of $\mathcal{W}$ occurs at the critical point $r_{0}=1$. This fact proves the critical behavior of the non-Hermitian systems is faithfully detected and witnessed by the EUR witness.
\begin{figure}[htpb]
  \begin{center}
     \includegraphics[width=7.5cm]{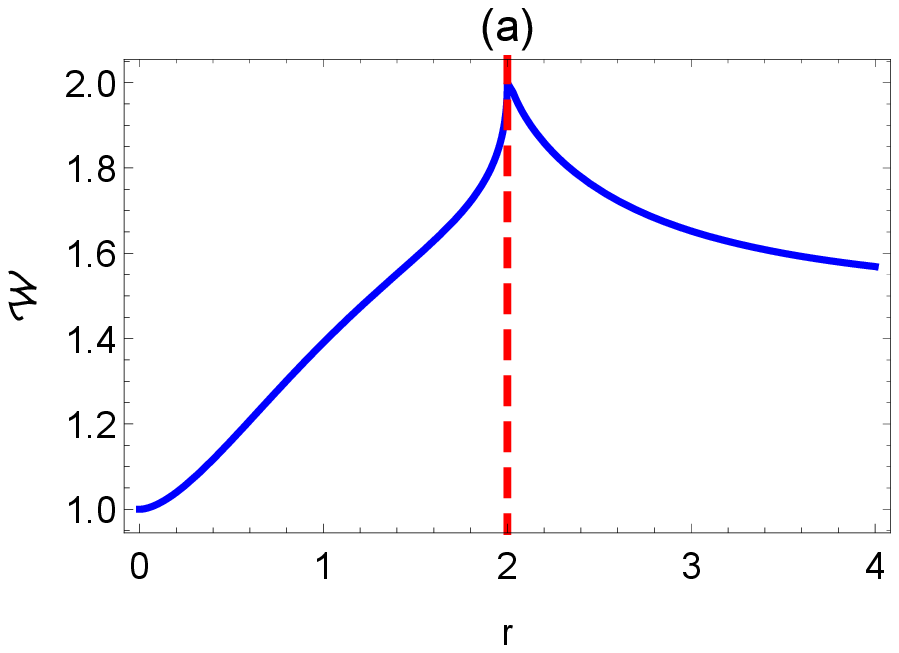}
     \includegraphics[width=7.5cm]{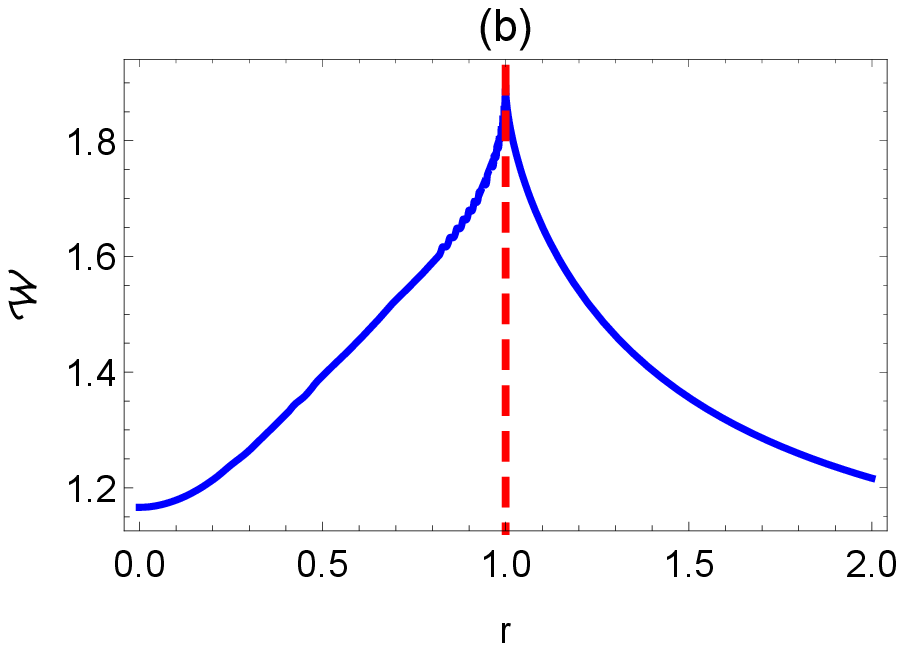}
     \caption{\label{fig2} (Color online) (a) The criticality witness $\mathcal{W}$ in a $PT$ symmetric non-Hermitian system with $\sigma=s=2$ and $\phi=\pi/2$. Therefore, a quantum critical point occurs at $r_{0}=2$. (b) The criticality witness $\mathcal{W}$ in a general non-Hermitian system with $\sigma=\sqrt{2}$, $s=\sqrt{2}/2$ and $\phi=\pi/2$. Therefore, there is a quantum critical point at $r_{0}=1$.}
  \end{center}
\end{figure}

To find a deeper insight into the nature of EUR with the critical point in the
non-Hermitian system, we have also investigated the rate of change of EUR in dynamics limit which can be realized experimentally and which can detect the exceptional point of non-Hermitian systems defined as
\begin{equation}\label{eq:H13}
	\mathcal{\beta} =\lim_{t\rightarrow\infty}\left\{\frac{d}{dt} [H(R)+H(Q)]\right\}.
	\end{equation}
The non-vanishing parameter $\beta$ is corresponding to the unbroken phase regime, while the vanishing of $\beta$ is corresponding to the broken phase regime. Hence $\beta$ is similar to the local order parameter characterizing quantum phase transitions. Results are presented in Fig. 3 where a sudden change of $\beta(r)$ occurs at critical point, above which the non-vanishing value of $\beta(r)$ is observed in the unbroken phase regime and below which the value of $\beta(r)$ vanishes in the broken phase regime. Therefore, $\beta(r)$ can be used as an efficient tool to detect and characterize the phase transition which can be realized experimentally.

\begin{figure}[htpb]
  \begin{center}
     \includegraphics[width=7.5cm]{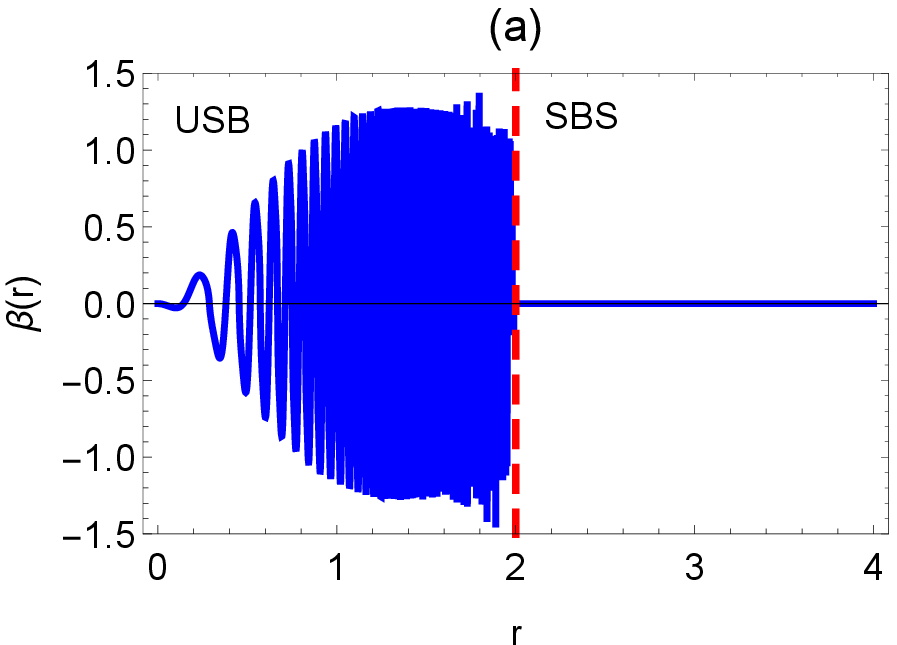}
     \includegraphics[width=7.5cm]{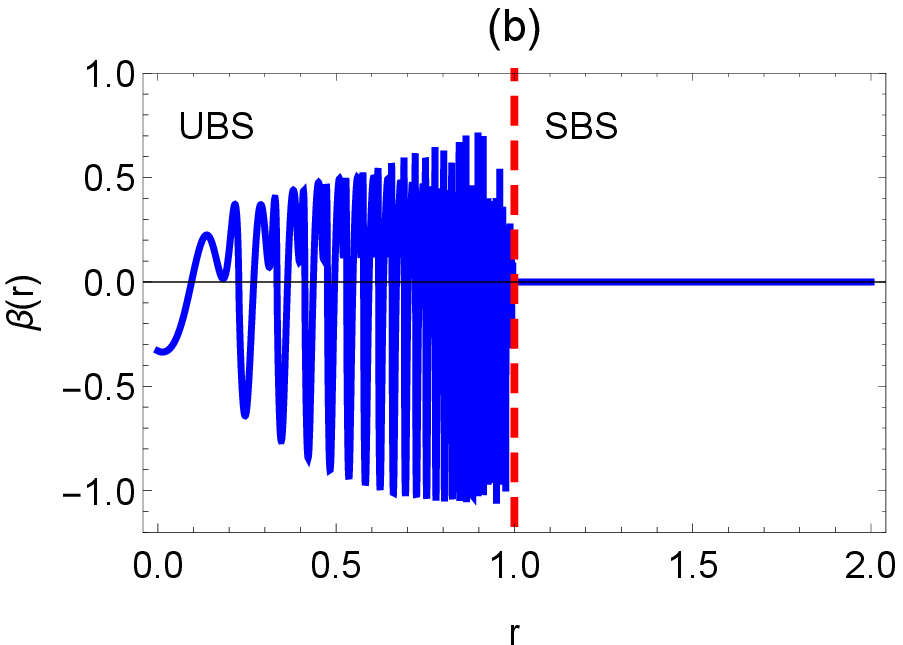}
     \caption{\label{fig3} (Color online) (a) The rate of change of EUR $\beta(r)$ in a $PT$ symmetric non-Hermitian system with $\sigma=s=2$ and $\phi=\pi/2$. Therefore, a quantum critical point occurs at $r=2$. (b) The rate of change of EUR $\beta(r)$ in a general non-Hermitian system with $\sigma=\sqrt{2}$, $s=\sqrt{2}/2$ and $\phi=\pi/2$. Therefore, there is a quantum critical point at $r=1$.}
  \end{center}
\end{figure}

\subsection{Criticality witness in anti-PT-symmetric non-Hermitian systems}
The generalized form of a single-qubit anti-PT symmetric Hamiltonian can be expressed as~\cite{Long}
\begin{eqnarray}
	\mathcal{H}_{NH} =\begin{pmatrix}
	             \lambda e^{i\phi}          &         is   \\
	             i s           &        -\lambda e^{-i\phi}
	              \end{pmatrix},
	\end{eqnarray}
where both of the parameters $\lambda$ and s denote real numbers.  It is easy to show that this Hamiltonian satisfies the anti PT invariant Hamiltonian $(PT)H(PT)=-H$~\cite{Benderc,Peng}. The anti-PT symmetric non-Hermitian Hamiltonian has complex eigenvalues spectra $ \varepsilon_{\pm} = i \lambda\sin\phi\pm \sqrt{\lambda^2 \cos^2\phi-s^2}$. Therefore, the system is termed in the regime of unbroken phase when $|s| >| \lambda \cos\phi|$~\cite{Long}. Following the same calculation procedure as above, the analytical expression of EUR is obtained

\begin{equation}\label{Eq14}
EUR=-\sum_{i=1}^{2}p_{x}^{i}\log_{2}{p_{x}^{i}}-\sum_{i=1}^{2}p_{z}^{i}\log_{2}{p_{z}^{i}}
\end{equation}
with

$p_{x}^{1}=\frac{1}{2}(1+\frac{s\omega\sinh2\omega t}{s^2\cosh2\omega t-\lambda^2\cos^2\phi})$,

$p_{x}^{2}=\frac{1}{2}(1-\frac{s\omega\sinh2\omega t}{s^2\cosh2\omega t-\lambda^2\cos^2\phi})$,

$p_{z}^{1}=\frac{1}{2}(1-\frac{\omega^2}{s^2\cosh2\omega t-\lambda^2\cos^2\phi})$,

$p_{z}^{2}=\frac{1}{2}(1+\frac{\omega^2}{s^2\cosh2\omega t-\lambda^2\cos^2\phi})$.

with $\omega=\sqrt{s^2-\lambda^2 \cos^2\phi}$

\begin{figure}[htpb]
  \begin{center}
     \includegraphics[width=7.5cm]{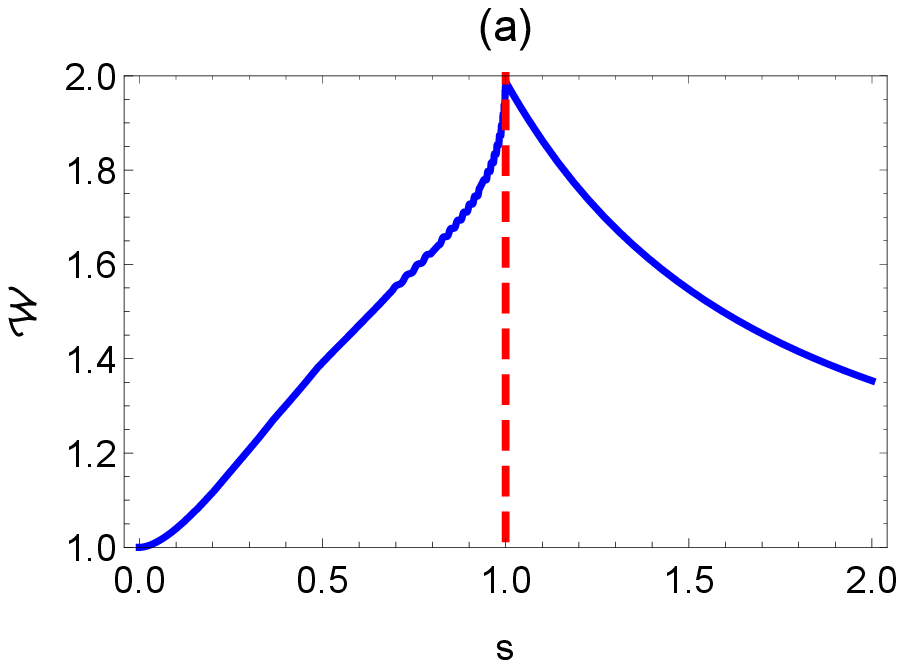}
      \includegraphics[width=7.5cm]{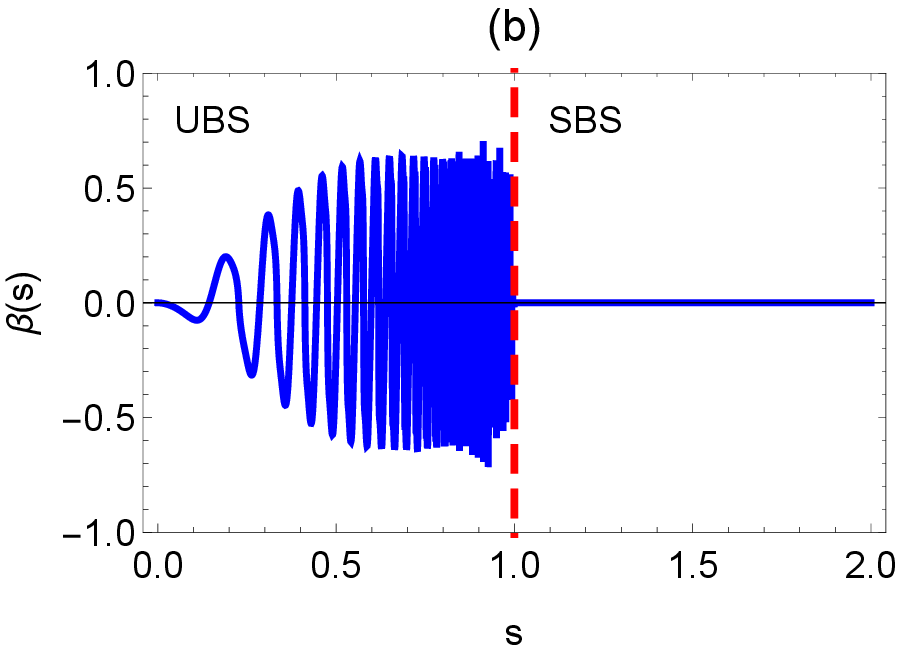}
     \caption{\label{fig4} (Color online) (a) The rate of change of entropic uncertainty relation $\beta(s)$ in a anti-$PT$ symmetric non-Hermitian system. (b) The criticality witness of the the anti-parity-time symmetric non-Hermitian systems for initial state $|\Psi\rangle=\frac{1}{\sqrt{2}}(|0\rangle+|1\rangle)$. Other parameters are fixed $\lambda=1$ and $\phi=\pi/2$.}
  \end{center}
\end{figure}

As expected, the efficiency of the proposed criticality witness also works for anti-PT-symmetric non-Hermitian system where we take $\lambda=1$ and $\phi=\pi/2$.  A abrupt change of $\mathcal{W}$ happens at the transition point $s_{0}=\lambda \cos\phi=1$, as reported in Fig. 4(a). This indicates the critical behavior of the anti-parity-time symmetric non-Hermitian systems is also identified by criticality witness $\mathcal{W}$. On the other hand,
the revival of $\beta(s)$ for an anti-parity-time symmetric non-Hermitian system is also observed in Fig. 4(b). The non-vanishing parameter $\beta(s)$ is corresponding to the broken phase regime, while the vanishing of $\beta(s)$ is corresponding to the unbroken phase regime. The signature of a quantum
critical point is characterized by $\beta(s)$ from non-zero value to zero value.

\subsection{Experimental feasibility}
Finally, we give a brief discussion of the above predictions which are easy to implement in experiment according to our procedure. To accomplish this, we here restrict our discussions to the single-ion system where the PT-symmetric Hamiltonian with balanced gain and loss has been realized in experiment~\cite{Chenpx}. The experimental setup refers to the trapped $40Ca^{+}$ ion in a magnetic field and be chosen four Zeeman energy levels labeled by $||0\rangle = |S_{-1/2}\rangle$, $|1\rangle = |D_{5/2}\rangle$, $|2\rangle=|S_{1/2}\rangle$ and $|3\rangle=|P_{3/2}\rangle$. First of all, the ion is initially prepared in the ground state $|0\rangle$, which is driven to the excited state $|1\rangle$ by the first laser. At the same time, another laser is switched on between $|1\rangle $ and $|3\rangle$ which decays quickly to the state $|2\rangle $. Therefore, the loss rate $\gamma$ between $|1\rangle $ and $|2\rangle$ can be realized by adjusting the intensity of the second laser beam. Under this condition, an effective PT-symmetric Hamiltonian $H_{eff}=\frac{\Omega}{2}\sigma_{x}-i\frac{\gamma}{2}\sigma_{z}$ is obtained~\cite{Chenpx}, here $\Omega$ is the coupling rate between $|0\rangle $ and $|1\rangle$. Compared with the Eq.(3), one should take $\sigma$=$s$=$\Omega/2$, $r$=$-\gamma/2$ and $\phi=\pi/2$. Therefore, to verify the EUR dynamical features as well as criticality witness, the probabilities $p_{i}=Tr(\sigma_{i}\rho)$ should be measured from density-matrix of $\rho$ due to the fact that the density-matrix elements of $\rho$ can be directly measured in experiment~\cite{Chenpx}.

\section{Conclusion} \label{conclusions}

In conclusion, we have investigated the dynamics of EUR in the non-Hermitian systems. Compared with the dynamics governed by a Hermitian Hamiltonian, there are three different types of behavior depending on the dynamics of EUR in the unbroken regime, at the exceptional point, or in the spontaneously broken regime. In the unbroken regime, the EUR undergoes an oscillatory behavior, while in broken phase regime where the oscillation breaks down. At the exceptional point, the EUR increases asymptotically to a stable value. The exceptional point marks the oscillatory and non-oscillatory behavior of the EUR. In addition, we also identify the witness of critical behavior in terms of the EUR in the dynamical limit where two kinds of non-Hermitian models, including (anti-)PT-symmetric systems are taken into consideration. We find that criticality witness can be an effective index to identify exceptional point in these two models. Finally, the experimental feasibility of our schemes is also discussed.

Our approach establishes a general connection between the criticality of non-Hermitian system and the behaviors of EUR. Therefore, our results may have potential applications to witness and detect criticality in non-Hermitian systems. Besides, our investigation could be helpful for understanding the dynamics features of EUR and further manipulating the EUR in the non-Hermitian system.

\section{Acknowledgments}
The authors thank Maofa Fang and Pingxing Chen for kindly help. This work is supported by by the National Natural Science Foundation of China (Grant No.11747107), the Natural Science Foundation of Hunan Province (Grant No.2021JJ30757), the Scientific Research Project of Hunan Province Department of Education (Grant Nos.19B060 and 19C0539). Y N Guo is supported by Training Program for Excellent Young Innovators of Changsha(kq2009076).

\label{app:eff-trans}






\end{document}